\documentclass[twocolumn,prl,showpacs,superscriptaddress,preprintnumbers,amsmath,amssymb,floatfix]{revtex4-2}
\usepackage{graphicx}

\usepackage[normalem]{ulem}

\usepackage{braket}

\usepackage[]{color}

\begin{document}
%\preprint{APS/123-QED}

%Title of paper
\title{Selective orbital imaging of excited states with x-ray spectroscopy: the example of $\alpha$-MnS}

\author{A.~Amorese}
    \affiliation{Max Planck Institute for Chemical Physics of Solids, N{\"o}thnitzer Str. 40, 01187 Dresden, Germany}
    \affiliation{Institute of Physics II, University of Cologne, Z\"{u}lpicher Str. 77, D-50937 Cologne, Germany}

\author{B.~Leedahl}
    \affiliation{Max Planck Institute for Chemical Physics of Solids, N{\"o}thnitzer Str. 40, 01187 Dresden, Germany}

\author{M.~Sundermann}
    \affiliation{Max Planck Institute for Chemical Physics of Solids, N{\"o}thnitzer Str. 40, 01187 Dresden, Germany}
    \affiliation{Institute of Physics II, University of Cologne, Z\"{u}lpicher Str. 77, D-50937 Cologne, Germany}

\author{H.~Gretarsson}
    \affiliation{Max Planck Institute for Chemical Physics of Solids, N{\"o}thnitzer Str. 40, 01187 Dresden, Germany}
    \affiliation{PETRA III, Deutsches Elektronen-Synchrotron (DESY), Notkestra{\ss}e 85, 22607 Hamburg, Germany}

\author{Z.~Hu}
    \affiliation{Max Planck Institute for Chemical Physics of Solids, N{\"o}thnitzer Str. 40, 01187 Dresden, Germany}

\author{H.-J.~Lin}
    \affiliation{National Synchrotron Radiation Research Center, 101 Hsin-Ann Road, Hsinchu 30076, Taiwan}

\author{C.~T.~Chen}
    \affiliation{National Synchrotron Radiation Research Center, 101 Hsin-Ann Road, Hsinchu 30076, Taiwan}

\author{M.~Schmidt}
    \affiliation{Max Planck Institute for Chemical Physics of Solids, N{\"o}thnitzer Str. 40, 01187 Dresden, Germany}
    
\author{H.~Borrmann}
    \affiliation{Max Planck Institute for Chemical Physics of Solids, N{\"o}thnitzer Str. 40, 01187 Dresden, Germany}

\author{Yu.~Grin}
    \affiliation{Max Planck Institute for Chemical Physics of Solids, N{\"o}thnitzer Str. 40, 01187 Dresden, Germany}

\author{A.~Severing}
    \affiliation{Max Planck Institute for Chemical Physics of Solids, N{\"o}thnitzer Str. 40, 01187 Dresden, Germany}
    \affiliation{Institute of Physics II, University of Cologne, Z\"{u}lpicher Str. 77, D-50937 Cologne, Germany}

\author{M.~W.~Haverkort}
    \affiliation{Institute for Theoretical Physics, Heidelberg University, Philosophenweg 19, 69120 Heidelberg, Germany}

\author{L.~H.~Tjeng}
    \affiliation{Max Planck Institute for Chemical Physics of Solids, N{\"o}thnitzer Str. 40, 01187 Dresden, Germany}

\date{\today}

\begin{abstract}
Herein we show that non-resonant inelastic x-ray scattering involving an $s$ core level is a powerful spectroscopic method to characterize the excited states of transition metal compounds. The spherical charge distribution of the $s$ core hole allows the orientational dependence of the intensities of the various spectral features to produce a spatial charge image of the associated multiplet states in a straightforward manner, thereby facilitating the identification of their orbital character. In addition, the $s$ core hole does not add an extra orbital angular momentum component to the multiplet structure so that  the well-established Sugano-Tanabe-Kamimura diagrams can be used for the analysis of the spectra. For $\alpha$-MnS we observe the spherical charge density corresponding to its high spin $3d^5$ ($^6A_1$) ground state configuration and we were able to selectively image its excited states and identify them as 
$t_{2g}$ ($^5T_2$) and $e_g$ ($^5E$) with an energy splitting $10Dq$ of 0.78\,eV.
\end{abstract}


\maketitle

Transition metal (TM) compounds display a wide variety of extraordinary properties of both theoretical and technological interest. These include metal-insulator and spin-state transitions, colossal magnetoresistance, various forms of magnetism and multiferroicity, as well as superconductivity. It is generally accepted that this richness in phenomena must somehow be related to the wealth of possible electronic states created by the strong Coulomb and exchange interactions between the TM valence electrons in the $d$ shell and the intricate interplay with the band formation in the solid \cite{Khomskii2014}. Understanding how those states are formed and which ones participate in the formation of the ground state and low lying excited states is a difficult task. It is therefore highly desirable if one can at least identify the relevant local charge, spin, and orbital degrees of freedom. To this end, input from experiments is necessary. While information about the valence and magnetic states can be routinely gathered using x-ray and neutron scattering techniques, the case for the orbital state is much more delicate since spectroscopic methods have to be applied and this often requires highly complex calculations to interpret the spectral lineshapes.

Recently, we have shown that non-resonant inelastic x-ray scattering (NIXS, also known as x-ray Raman scattering) involving an $s$ core hole ($s$-NIXS) is an experimental method that can provide a direct image of the local $d$ hole density in transition metal oxide single crystals \cite{Yavas2019,Leedahl2019}, i.e. the ground state $d$ orbital can be determined without the need for calculations to interpret the spectral lineshape. This opens up new opportunities for the investigation of the ground state, especially for those transition metal compounds that are too complex to be handled by \textit{ab-initio} theories. Here we will go one step further and explore the spectroscopy aspect of $s$-NIXS in order to study the excited states which are most often dominated by many-body atomic multiplet interactions. In particular we aim to determine the orbital character of those states. Our idea is that the use of an $s$ core hole should simplify the analysis of the spectra in two significant aspects. First, the presence of the $s$ hole does not add an extra orbital angular momentum so that use can be made of the well-established and readily available Sugano-Tanabe-Kamimura diagrams that depict the multiplet energy scheme of $3d$ ions for varying values of the crystal field \cite{Sugano1970}. Second, the identification of the orbital character of the multiplet state can be done in a direct manner by the imaging ability of the $s$-NIXS method \cite{Yavas2019,Leedahl2019}. To exemplify these points, we have carried out the experiment on $\alpha$-MnS, a rock salt type antiferromagnetic insulator with far from complete filling of the $d$ shell so that orbital degrees of freedom are present in its excited states.

\begin{figure}
    \centering
    \includegraphics[width=\columnwidth]{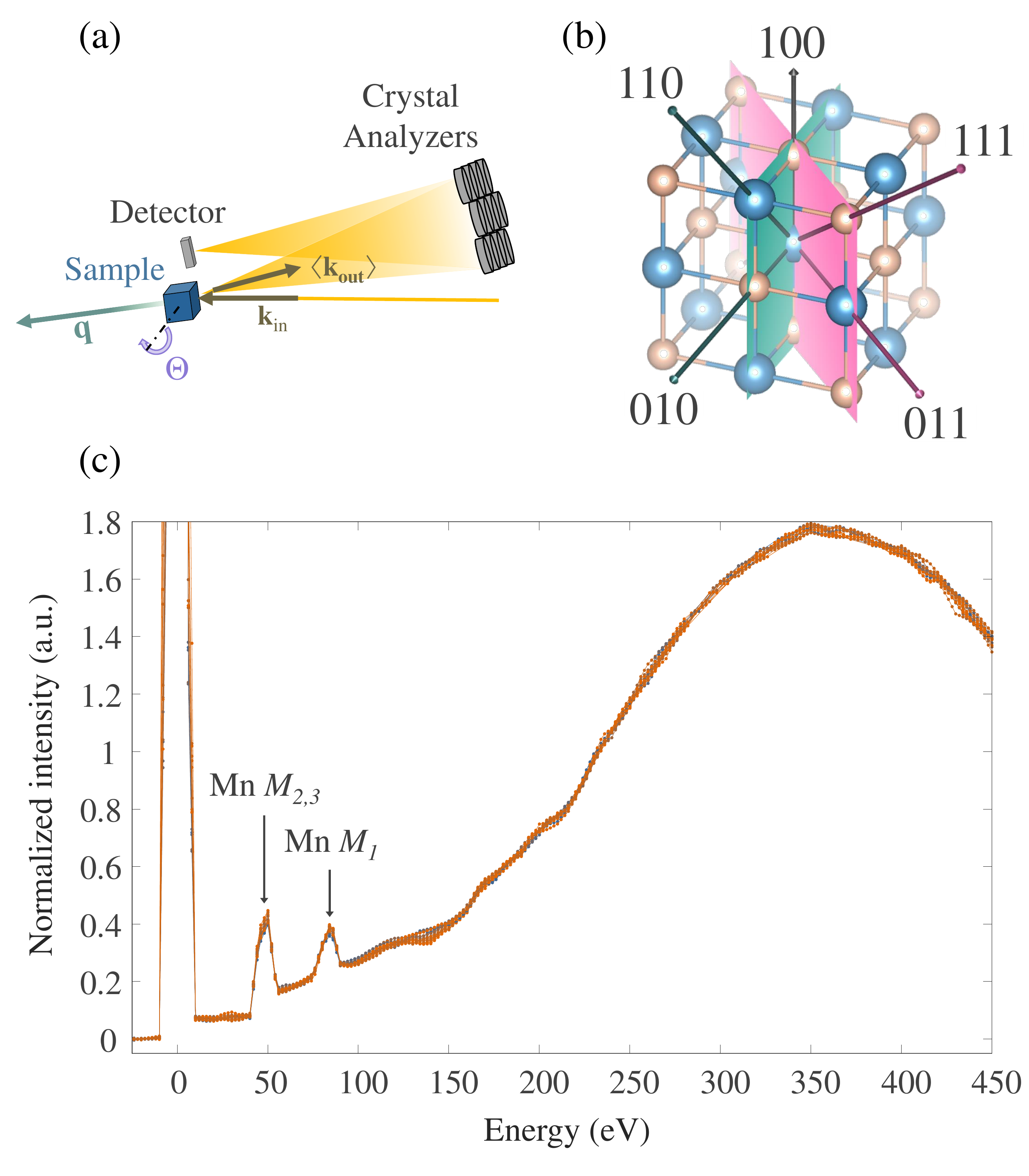}
    \caption{(a) Schematic representation of the NIXS spectrometer of P01 (DESY). During the experiment the angle $\Theta$ was varied to collect spectra at several directions within the same plane. $\Theta=0^\circ$ for $\mathbf{q}\parallel 100$, where $\mathbf{q}=\mathbf{k}_{in}-\mathbf{k}_{out}$ is the momentum transferred.
(b) Crystal structure of $\alpha$-MnS. The two planes explored are indicated in dark cyan and magenta. (c) Experimental spectra  acquired for various directions, normalized to the Compton profile which peaks at around 350 eV. The labels indicate the atomic-like Mn $M_{2,3}$ ($3p \rightarrow 3d$) and $M_1$ ($3s \rightarrow 3d$) transitions.}
\label{Fig1}
\end{figure}

The spectra were acquired on the NIXS endstation of the P01 beamline at PETRA III (DESY). 
The spectrometer (sketched in Fig.\,\ref{Fig1} (a) uses a Rowland geometry and consists of a $3\times  4$ array of 1\,m radius spherically bent Si crystal analyzers, which select the 9.7\,keV photons scattered from the sample using the Si(660) Bragg reflection and focus them on a 55\,$\mu$m-pixel Si Medipix3 2D detector \cite{Yavas2019}. During the acquisition of the spectra, the energy of the incident photons, monochromatized by a Si(311) double crystal monochromator, is continuously swept from $\approx$ 9.7\,keV (elastic line) towards higher energies, thus scanning the energy transferred in the inelastic scattering process. The experimental resolution, estimated from the full width at half maximum of the elastic line, was about 760\,meV. The scattering angle was set to $2\theta=155^\circ$ which yielded $\vert \mathbf{q} \vert\approx 9.6$\,\r{A}$^{-1}$, where $\mathbf{q}=\mathbf{k}_{in}-\mathbf{k}_{out}$ is the momentum transferred. 
The sample was a single crystal of rock salt type $\alpha$-MnS grown by chemical vapor transport (see Appendix). During the NIXS measurement the sample was kept at a temperature of 50\,K in order to prevent possible radiation damage. The spectra have been acquired at many $\Theta$ angles ($\Theta$ being zero when $\mathbf{q}\parallel [100]$) for two different orientations of the sample relative to the scattering plane, so that $\mathbf{q}\parallel$ [100], [111] and [011]  could be reached with the first orientation (magenta plane in Fig.\,\ref{Fig1}b) and $\mathbf{q}\parallel$ [100], [110] and [010] with the second orientation (dark cyan).
To reach the two orientations, the crystal was rotated by inserting the sample holder (a stainless steel pin, see Fig.\,\ref{Fig8}) respectively in the $0^\circ$ and $45^\circ$ slots available in the P01 cryostat. The angle between the two slots corresponds to a $45^\circ$ rotation of the crystal around its $[100]$ axis (which remains parallel to the momentum transfer $\mathbf{q}$ at $\Theta=0^\circ$) and the two resulting geometries allow to span the two planes of Fig.\,\ref{Fig1} (b) by varying the angle $\Theta$ as shown Fig.\,\ref{Fig1} (a).

Figure\,\ref{Fig1} (c) displays a compilation of the NIXS spectra measured for the different sample angles. They are characterized by a broad Compton profile peaking at 350\,eV, which was used to normalize the spectra and correct for the intensity variations due to self-absorption effects at different scattering geometries \cite{Yavas2019}. On top of this signal, two sharp peaks are visible, ascribable to atomic-like excitations, namely the Mn $M_{2,3}$ edge ($3p \rightarrow 3d$) at about 50\,eV and, relevant to this study, the dipole-forbidden $M_1$ ($3s \rightarrow 3d$) transition at around 83\,eV. Such a transition is well visible thanks to the large momentum transfer $\vert \mathbf{q} \vert$ \cite{Gordon2007,Bradley2010,Caciuffo2010,Willers2012}, which allows higher multipole excitations to gain intensity and become of use for quantitative analysis\cite{Yavas2019,Leedahl2019}.

\begin{figure}
    \centering
    \includegraphics[width=\columnwidth]{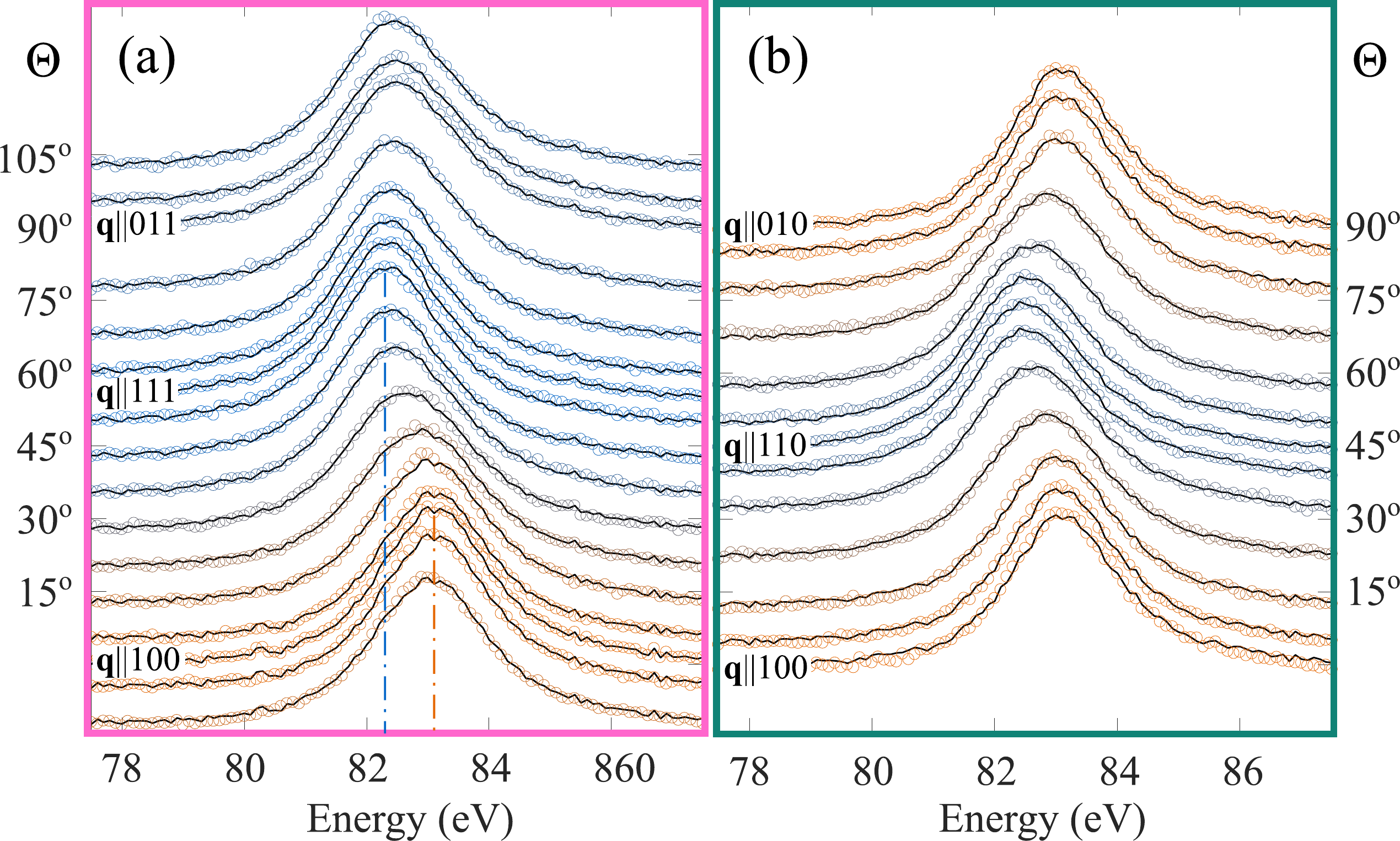}
    \caption{(a) and (b) Stack of the Mn $M_1$ NIXS spectra acquired for the planes
     as indicated in Fig.\,\ref{Fig1} (b). The black lines are the result of the fit procedure 
     explained later in the text. }
    \label{Fig2}
\end{figure}

The Mn $M_1$ edge spectra acquired at different $\Theta$ are shown in Fig.\,\ref{Fig2} (a) for the set $\mathbf{q}\parallel$ [100] --- [111] --- [011] (magenta) and in Fig.\,\ref{Fig2} (b) for $\mathbf{q}\parallel$ [100] --- [110] --- [010] (dark cyan). The Compton contribution has been subtracted from these spectra using a linear background.
At first glance, there are no prominent variations in the peak intensities, but, as highlighted by the colors, the overall peak energy position varies as a function of $\Theta$, i.e. as a function the direction of $\mathbf{q}$ with respect to the crystallographic axes.

\begin{figure*}
    \centering
    \includegraphics[width=\textwidth]{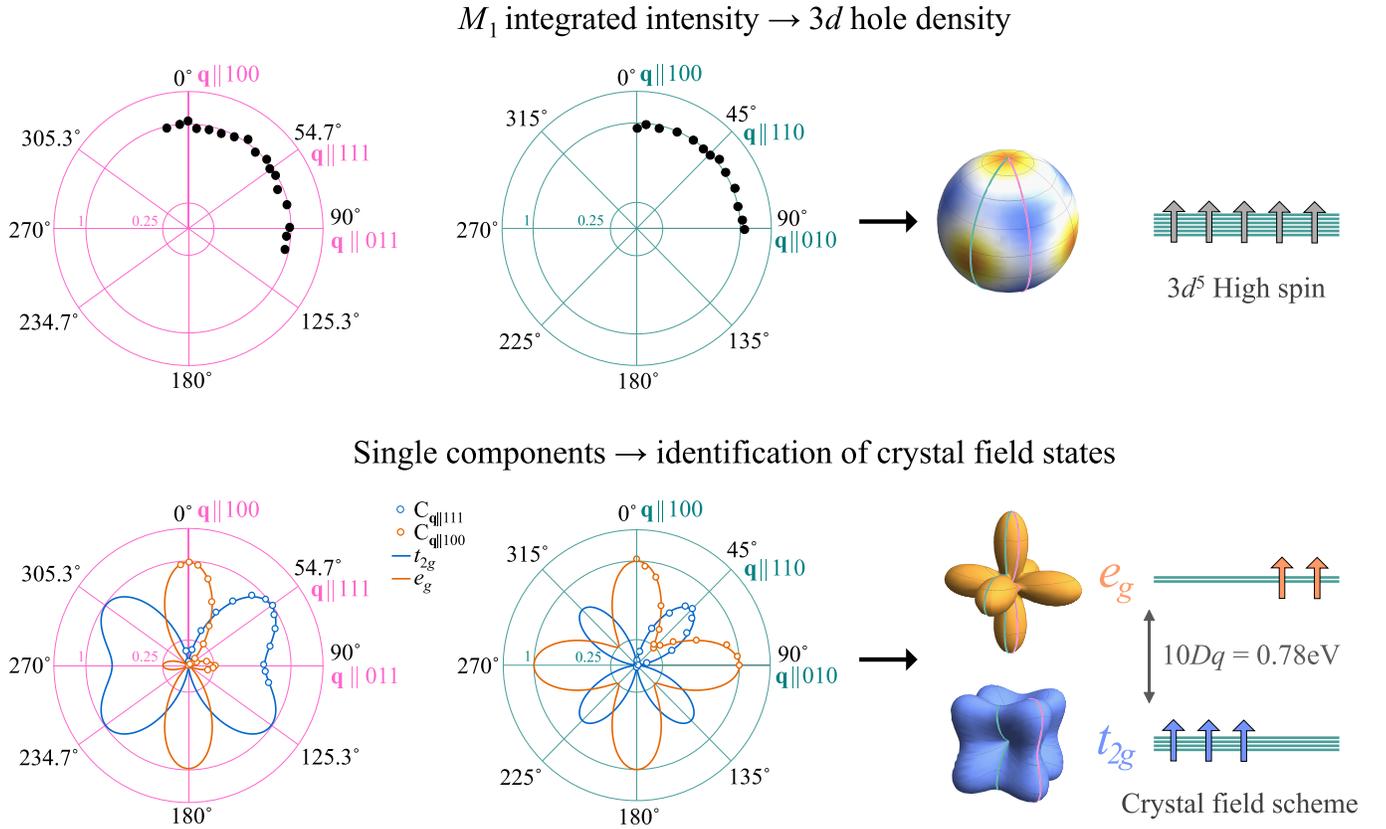}
    \caption{Top panel: Polar plot of the integrated intensity of the Mn $M_1$ in the [100]–[111]-[011] plane (far left, magenta) and in the [100]–[110]-[010] plane (middle left, dark cyan), together with the circular shaped projections of the spherical three-dimensional hole density (middle right) of the Mn high-spin $3d^5$ configuration (far right). Bottom panel: The intensity of the $\mathbf{q}\parallel$ [100] (orange) and $\mathbf{q}\parallel$ [111] (blue) components follow the angular dependence of the $e_g$ (orange) and $t_{2g}$ (blue) orbitals, respectively. The $e_g$/$t_{2g}$ energy splitting is about $10Dq$ = 0.78 eV.}
    \label{Fig3}
\end{figure*}

To quantitatively discuss our data, we display the integrated intensities of each  Mn $M_1$ spectrum of Fig.\,\ref{Fig2} on a polar plot (Fig.\,\ref{Fig3}, top panel). We can directly observe that there is no directional dependence, i.e. the integrated $M_1$ intensity is constant for all directions. As explained earlier \cite{Yavas2019}, given that the core-level $s$ orbital is spherical, the directional dependence of the integrated $M_1$ intensity is only determined by (and directly proportional to) the density of $d$ holes in the direction parallel to $\mathbf{q}$. In other words, this directional dependence provides a direct spatial image of the local empty valence states. We thus observe from the polar plots in the top panel of Fig.\,\ref{Fig3} that the Mn $3d$ hole density is spherical. This is consistent with the scenario in which all five spin-up or all five spin-down $3d$ orbitals are unoccupied, i.e. in which the Mn$^{2+}$ $3d^5$ ion is in its Hund's rule high-spin $^6A_1$ ground state. It is important to note once again that this result does not rely on any modeling or calculations about the electronic structure of the compound, but that it is directly obtained simply by measuring the orientational dependence of the integrated intensities of the spectra.\\

While the integrated intensity of the spectra is constant with $\Theta$, the energy position of the $M_1$ peak in Fig.\,\ref{Fig2} (a) and (b) does vary with the direction, between 82.37\,eV (blue vertical line) for the $\mathbf{q}\parallel [111]$ spectrum and 83.15\,eV (orange vertical line) for $\mathbf{q}\parallel [100]$. Since no discernable dispersion can be expected for core-hole excitations, such a variation in energy indicates that the $M_1$ signal consists of features positionend at different energies whose relative intensities change with $\Theta$. With the Mn ion coordinated octahedrally by six S ions, we expect that the energy differences in the final states must be related to $10Dq$, the octahedral crystal field splitting between the $t_{2g}$ and $e_g$ orbitals. 

\begin{figure}
    \centering
    \includegraphics[width=1\columnwidth]{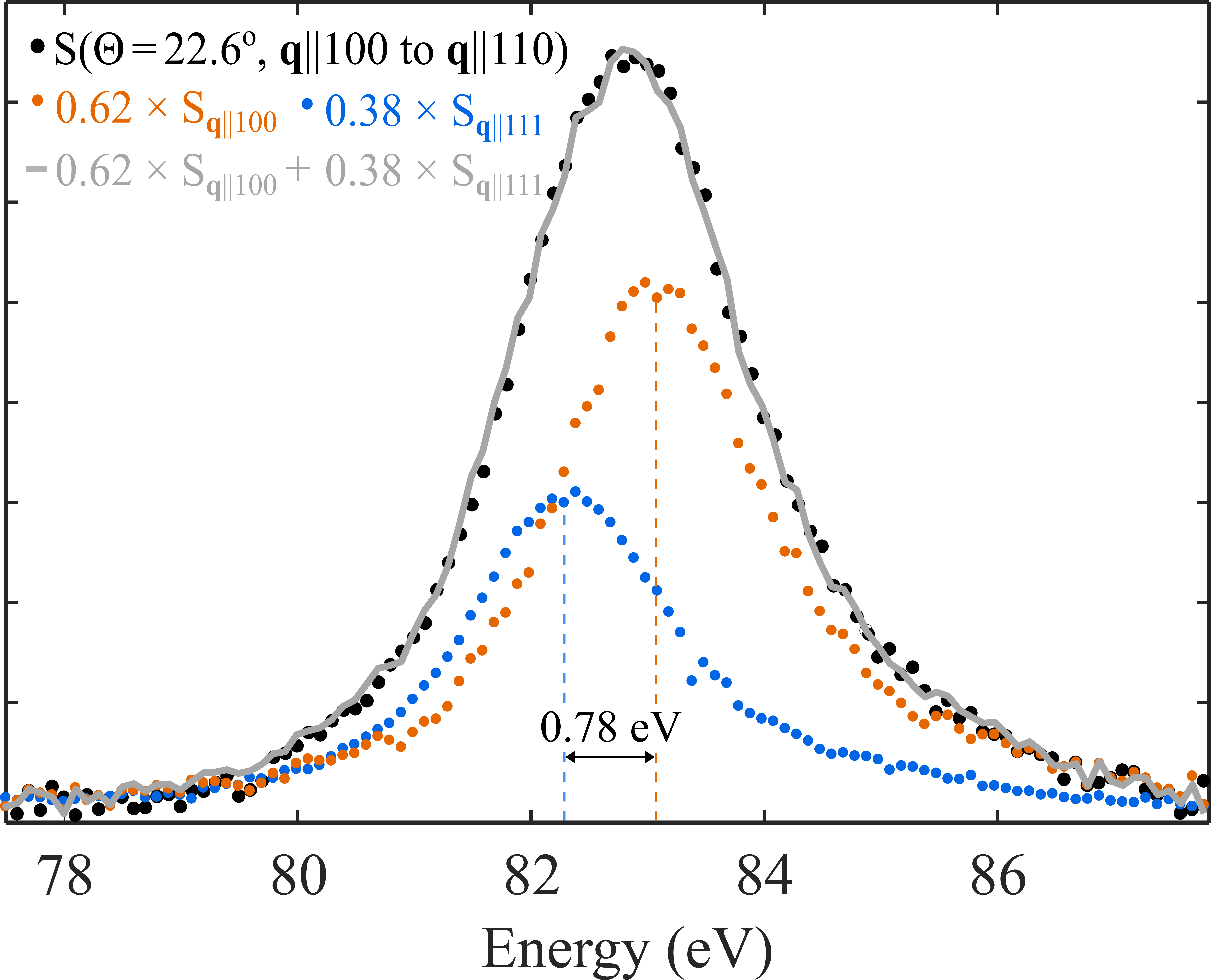}
    \caption{The experimental spectrum for a given angle $\Theta$ (black dots) is decomposed using the weighted sum of $S_{\mathbf{q}\parallel [111]}$ (blue dots) and $S_{\mathbf{q}\parallel [100]}$ (orange dots) spectra that provides the best fit (grey line).}
    \label{Fig4}
\end{figure}

We now make the \textit{ansatz} that the two spectra acquired at $\mathbf{q}\parallel [111]$ and $\mathbf{q}\parallel [100]$, which show the extreme peak positions and narrowest line shapes, are the basic components making up the $M_1$ NIXS signal for every other $\Theta$ value. We performed fits to all spectra using a linear combination of these two experimental spectra and determined their relative weights, so that each spectrum is described as $S(\Theta)=c_{111}(\Theta) S_{\mathbf{q}\parallel 111} + c_{100}(\Theta) S_{\mathbf{q}\parallel 100}$ where $c_{111}$ and $c_{100}$ are the free fitting parameters, as depicted in Fig.\,\ref{Fig4}. The resulting fits are shown with gray lines in Fig.\,\ref{Fig2} (a) and (b). The weights $c_{111}(\Theta)$ and $c_{100}(\Theta)$ obtained by the fits are plotted in Fig.\,\ref{Fig3} (bottom panel) for the two sample orientations. We see that the angular dependence follows the shape of the $t_{2g}$ and $e_g$ states with great accuracy, allowing us to directly identify the orbitals reached in each excitation. In particular, the $t_{2g}$ orbital shape is drawn by the angular dependence of $c_{111}(\Theta)$, the weight of the $S_{\mathbf{q}\parallel 111}$ component peaking at 82.37\,eV. Likewise, the excitation into $e_g$ orbitals is represented by $c_{100}(\Theta)$, the weight of the $S_{\mathbf{q}\parallel 100}$ component peaking at 83.15\,eV. The difference between these two energies is due to the $e_g$\,-\,$t_{2g}$  splitting, and it is therefore a direct measurement of the crystal field parameter $10Dq=83.15$\,eV$-82.37$\,eV$= 0.78$\,eV.

\begin{figure*}
    \centering
    \includegraphics[width=0.49\textwidth]{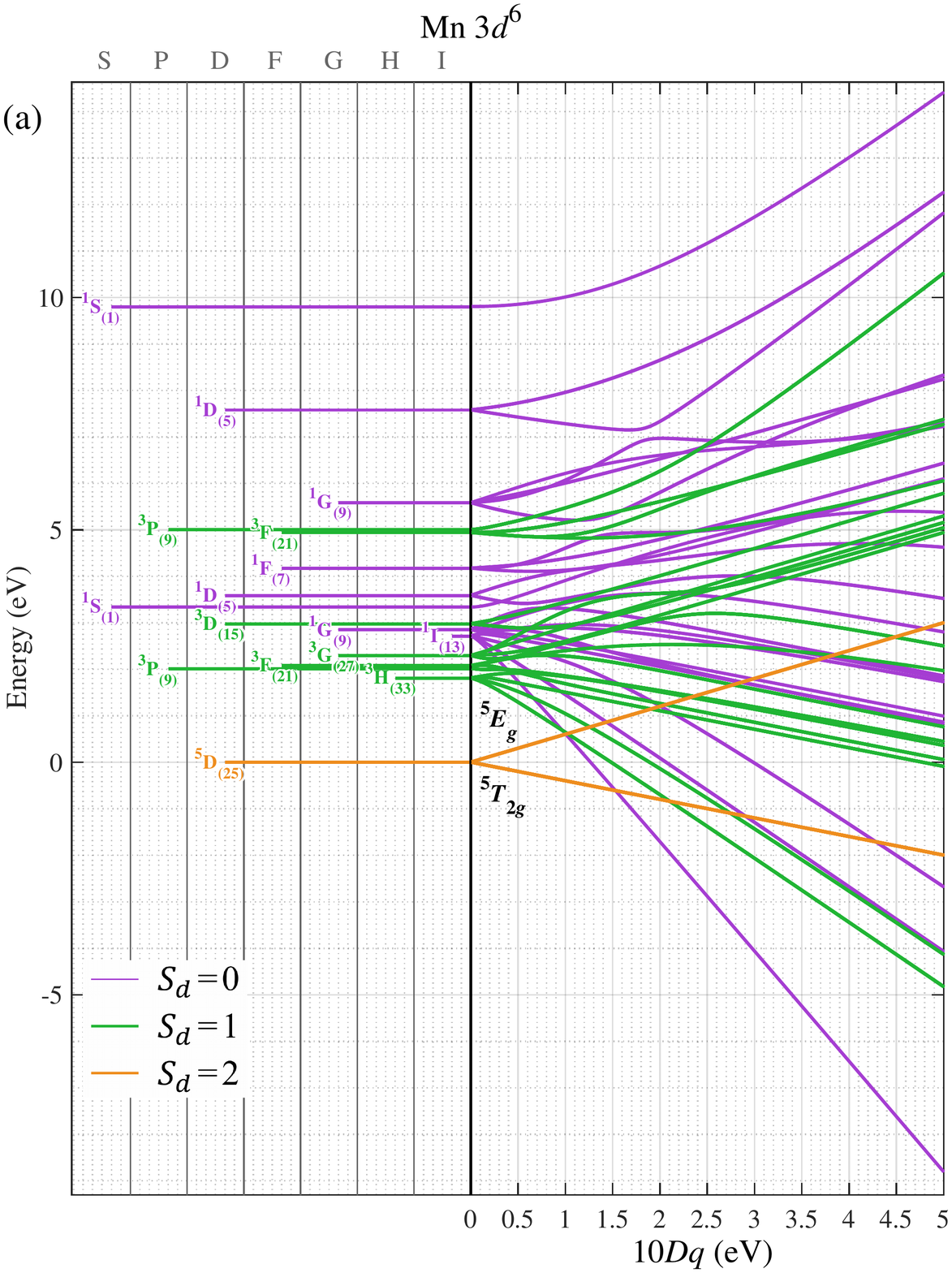}
    \includegraphics[width=0.49\textwidth]{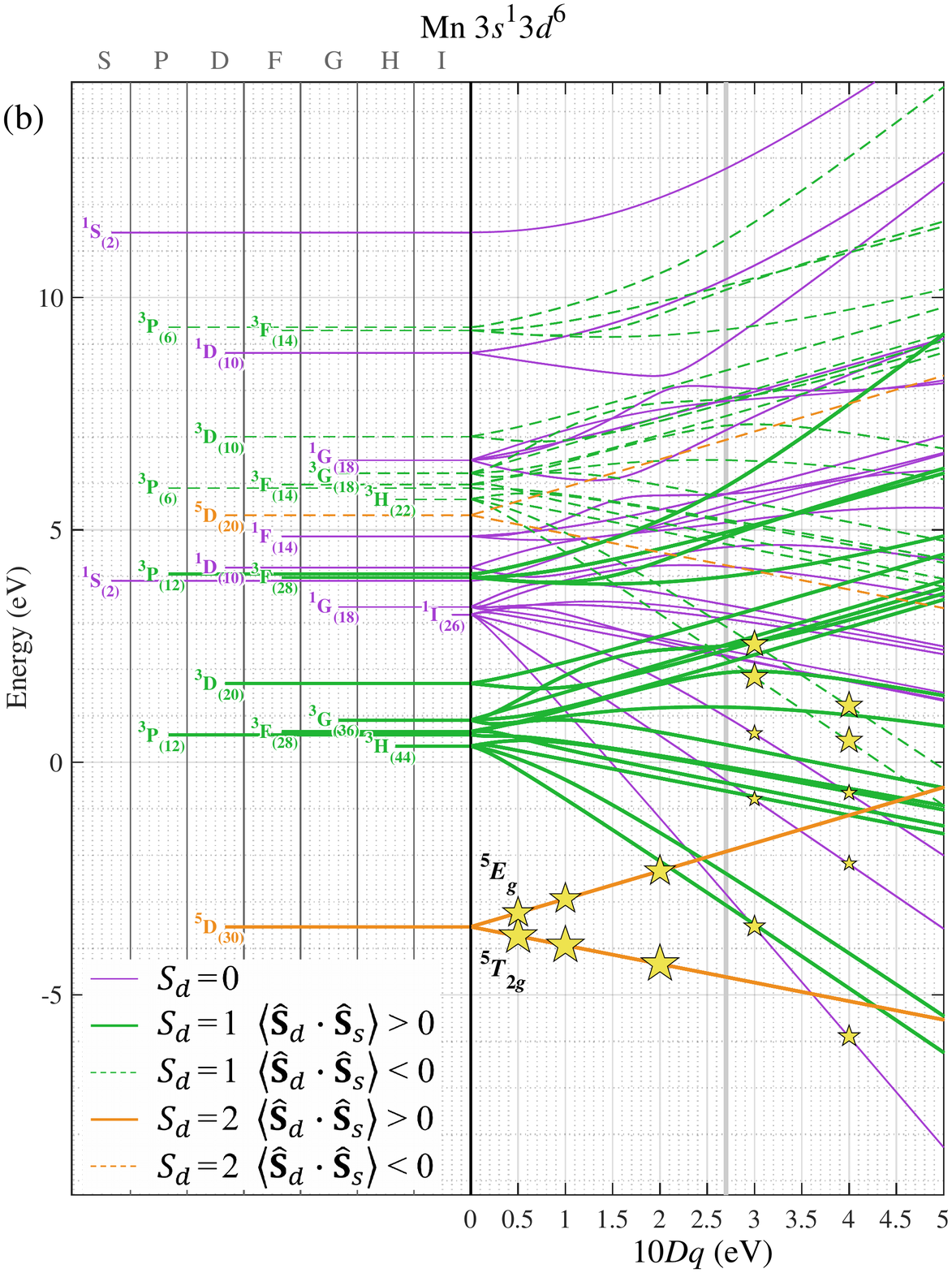}
\caption{Modified Sugano-Tanabe-Kamimura diagrams showing the dependence on $10Dq$ of the energy levels of (a) Mn $3d^6$ (in absence of the core $3s$ hole) and of (b) Mn $3s^13d^6$ (the $s$-NIXS final state). 
The 0\,eV energy reference is the same for both panels, and it is fixed to the lowest state of Mn $3d^6$ at $10Dq=0$\,eV. 
Each multiplet is labeled with the corresponding term symbol at $10Dq=0$\,eV  and in parenthesis it is indicated its degeneracy. The colors indicate each state's spin multiplicity: purple for singlets, green for triplets and orange for quintets. 
In panel (b), the thick full lines represent the states having parallel $\bold{S}_d$ and $\bold{S}_s$, the thin dashed lines the states with $\bold{S}_d$ anti-parallel to $\bold{S}_s$, and the thin full lines the states not affected by $\bold{S}_s$. The stars indicate, for several $10Dq$, the possible final states of the $s$-NIXS process, and the vertical gray line at $10Dq=2.7$\,eV is the transition between the high and low spin configurations of the $3d^5$ ground state.
The Slater parameters used for Mn $3d^6$, after reduction to 80\% of their atomic value, are: $F^2_{3d-3d}=7.258$\,eV and $F^4_{3d-3d}=4.472$\,eV. The reduced Slater parameters for Mn $3s^13d^6$ are: $F^2_{3d-3d}=8.426$\,eV, $F^4_{3d-3d}=5.244$\,eV, and $G^2_{3s-3d}=8.856$\,eV.}
\label{Fig5}
\end{figure*}

A proper interpretation of the spectra must include the effects of Coulomb and exchange interactions between the electrons within the $d$ shell, meaning that we need to put our results in a many body framework which takes into account both the full atomic multiplet theory and the local effect of the lattice. In particular, starting from the $(3s^2)3d^5$ configuration for the ground state of the Mn ion in $\alpha-$MnS, the $s$-NIXS process leads to a $3s^13d^6$ final state. Since the $s$ core hole does not add an extra orbital angular momentum component, the multiplet structure of the final state will be closely related to the one of the $3d^6$ configuration. 
Therefore, one could make use of the well-estabilished Sugano-Tanabe-Kamimura diagrams \cite{Sugano1970} for a quantitative analysis of $s$-NIXS spectra, after taking into account for some modifications due to the presence of the extra $3s$ spin. 
We exemplify this in Figs.\,\ref{Fig5} (a) and (b), where we reproduce the total energy diagrams for a Mn ion in $3d^6$ and $3d^13d^6$ configuration calculated with the \textit{Quanty} code \cite{Haverkort2012}.
The corresponding Slater integrals $F^2_{3d-3d}$ and $F^4_{3d-3d}$ (and $G^2_{3s-3d}$ for the $3s^13d^6$ configuration) have been obtained using the code by R. D. Cowan \cite{Cowan1981} and reduced to 80\% of their Hartree-Fock value to account for configuration interaction effects not included in the calculation \cite{deGroot1994,Tanaka1994}. A simplified version of these diagrams, containing only a selection of the multiplets to improve readability, is displayed in Fig.\,\ref{Fig9} and can be found together with a brief explanation in the Appendix.

To ease the comparison between the diagrams of the two configurations the lowest energy state is not fixed at 0\,eV for every $10Dq$. Instead, the lowest state of the $3d^6$ configuration is put to zero only for $10Dq=0$\,eV, and serves as reference energy. In this way, the multiplets are split by the action of $10Dq$, but the average energy of the diagram is kept constant.

The energy levels are labeled for zero crystal field ($10Dq=0$\,eV, spherical symmetry) with the $^{(2S_d+1)}L_d$ term symbols, where $S_d$ and $L_d$ represent the quantum numbers of the $3d$ shell and the orbital quantum numbers $L_d$ are indicated with the usual letter notation ($S$, $P$, $D$, $F$, $G$...). The subscript in parenthesis indicates the degeneracy of each term symbol, also including the degeneracy given by the $3s$ hole. In total, the $3d^6$ configuration has $\binom{10}{6} = 210$ states, while $3s^13d^6$ has $2 \times \binom{10}{6} = 420$ states, due to the extra multiplicity of the $3s^1$ spin. The colors group the states in the diagrams according to their $3d$ spin multiplicity ($2S_d+1$): singlets are purple, triplets green, and quintets orange.

Starting from the singlets, it is straightforward to notice that the energy scheme within the set of singlets, as well as the dependence of each state on $10Dq$, is essentially the same between the two configurations $3d^6$ and $3s^13d^6$. This agrees with the fact that singlet states cannot be modified by an interaction with a $3s$ spin, given by $\braket{\hat{\bold{S}}_d\cdot\hat{\bold{S}}_s}$, which is zero for $S_d=0$. The minor changes between the two groups ($\lesssim 10\%$ of the energy splittings) is due to the different values of the Slater integrals $F^2_{3d-3d}$ and $F^4_{3d-3d}$ between the two configurations. 

The sets of triplets (green lines) and quintets (orange) of the $3s^13d^6$ configuration, instead, are further divided in two subgroups depending on the relative alignment of $\bold{S}_d$ and $\bold{S}_s$. The states having the two spins parallel are represented with thick lines, while the thin dashed lines represent the states with opposite $\bold{S}_d$ and $\bold{S}_s$. These subgroups are each a replica of the corresponding $3d^6$ triplet or quintet set, rigidly shifted in energy due to the spin interaction. 

\begin{table}[]
\def\arraystretch{1.7}
\begin{tabular}{|c|c|c|c|c|c|}
\hline 
\rule[-1ex]{0pt}{2.5ex} & & \multicolumn{2}{c|}{$S_{\text{tot}}=S_d + S_s$} & \multicolumn{2}{c|}{$S_{\text{tot}}=S_d - S_s$} \\ 
\hline 
\rule[-1ex]{0pt}{2.5ex} $S_d$ & $2S_d+1$ & $\braket{\hat{\bold{S}}_d\cdot\hat{\bold{S}}_s}$ & $\Delta E$  [$G^2_{3s-3d}$] & $\braket{\hat{\bold{S}}_d\cdot\hat{\bold{S}}_s}$ & $\Delta E$ [$G^2_{3s-3d}$]  \\ 
\hline 
\rule[-1ex]{0pt}{2.5ex} 0             & 1 & $0$            & $0$           & $0$                & $0$ \\ 
\hline                                                                           
\rule[-1ex]{0pt}{2.5ex} $\frac{1}{2}$ & 2 & $\frac{1}{4}$  & $-\frac{1}{10}$ & $-\frac{3}{4}$   & $\frac{3}{10}$\\ 
\hline                                                                                                           
\rule[-1ex]{0pt}{2.5ex} 1             & 3 & $\frac{2}{4}$  & $-\frac{2}{10}$ & $-\frac{4}{4}$   & $\frac{4}{10}$\\ 
\hline                                                                                                           
\rule[-1ex]{0pt}{2.5ex} $\frac{3}{2}$ & 4 & $\frac{3}{4}$  & $-\frac{3}{10}$ & $-\frac{5}{4}$   & $\frac{5}{10}$\\ 
\hline                                                                                                           
\rule[-1ex]{0pt}{2.5ex} 2             & 5 & $\frac{4}{4}$  & $-\frac{4}{10}$ & $-\frac{6}{4}$   & $\frac{6}{10}$\\ 
\hline                                                                                                           
\rule[-1ex]{0pt}{2.5ex} $\frac{5}{2}$ & 6 & $\frac{5}{4}$  & $-\frac{5}{10}$ & $-\frac{7}{4}$   & $\frac{7}{10}$\\ 
\hline                                                    
\end{tabular} 
\caption{Eigenvalues of $\hat{\bold{S}}_d\cdot\hat{\bold{S}}_s$, related to the energy splitting due to the Coulomb interaction of pure spin multiplets (i.e. neglecting spin-orbit interaction) in the $d$ shell with the open $3s^1$ core shell for states with the $d$ spin either parallell ($S_{\text{tot}}=S_d + S_s$) or anti-parallell ($S_{\text{tot}}=S_d - S_s$) to the core $s$ spin.}\label{tab:Table1}
\end{table}

To understand and predict these energy shifts due to the Coulomb interaction between the $d$ electrons and the electron of the open $3s$ shell one can relate the Coulomb operator to the spin operators of the $d$ shell and $s$ shell. We have
\begin{equation*}
H_{3s-3d}^{Coulomb} = - \frac{2}{5} \hat{\bf{S}}_d \cdot \hat{\bf{S}}_s G^{2}_{3s-3d} + n_d( F^{0}_{3s-3d} - \frac{1}{10} G^{2}_{3s-3d} ).
\end{equation*}
The last term in the equation is constant for all multiplets within a $3s^1\,3d^{n_d}$ configuration and as such does not lead to a splitting between the different states. The eigenvalues of $\hat{\bold{S}}_{d}\cdot\hat{\bold{S}}_{s}$ can be obtained simply by inverting the formula\cite{Blundell} $ (\hat{\bold{S}}_\text{tot})^2=(\hat{\bold{S}}_{d})^2+(\hat{\bold{S}}_{s})^2+2\hat{\bold{S}}_{d}\cdot\hat{\bold{S}}_{s}$, where $S_\text{tot}$ can be $S_d + S_s = S_d +\frac{1}{2}$ or $\vert S_d - S_s \vert= \vert S_d -\frac{1}{2}\vert$, and remembering that the eigenvalue of $(\hat{\bold{S}})^2$ is $S(S+1)$. By applying these formulas, one obtains the eigenvalues listed in Table \ref{tab:Table1} for each possible value of $S_d$.\\

Typical values of $G^2_{3s-3d}$ for the $3d$ series, after a reduction to 80\,\% of their Hartree-Fock values, range from 8.5\,eV to 10.2\,eV. In general, $\Delta E$ is larger for larger spin multiplicities. With these ingredients, one can easily build the Sugano-Tanabe-Kamimura diagrams of the NIXS final configuration from the ones without the $3s$ core hole, listed, for example, in Fig. 5.1-5.7 of the famous book of S. Sugano, Y. Tanabe and H. Kamimura \cite{Sugano1970}. This allows the diagram of the possible $s$-NIXS final states to be reproduced without the need for performing new calculations and, by comparing the diagrams to the spectra, quantitatively determine the value $10Dq$.

The next step towards a complete understanding of the $s$-NIXS spectra is to realize that not all states depicted in Fig.\ref{Fig5} (b) can be reached starting from the ground state of Mn$^{2+}$ in $\alpha-$MnS. In our case, the addition of one extra $3d$ electron ($s = 1/2$) to the high-spin $^6A_1$ ($S_d = 5/2$) ground state can only lead to quintet final states ($S_d=2$), with the $3s^1$ spin parallel to the majority spin of the $3d$. Figure\,\ref{Fig10} in the Appendix describes this excitation process in more detail. Therefore, $-\frac{2}{5} G^2_{3s-3d} \braket{\hat{\bold{S}}_d\cdot\hat{\bold{S}}_s}<0$, and the low energy replica of the quintet set is reached. The possible $s$-NIXS final states for different values of $10Dq$ are indicated with stars in Fig.\,\ref{Fig5} (b), where the values of $10Dq$ corresponding to the initial $3d^5$ high-spin configuration $^6A_1$ are to the left of the gray vertical line, and low-spin on the right.  The size of each star is proportional to the intensity of the excitation, averaged over all directions. There are only two states in the diagrams that can be reached with a $s$-NIXS excitation, namely the $^5T_2$ (the extra $3d$ electron occupying $t_{2g}$ orbitals) and $^5E$ (the extra electron in the $e_g$). From the shape of the final state orbitals as imaged in Fig.\,\ref{Fig3} (bottom panels), we can identify immediately that the lower energy peak belongs to the $^5T_2$ state and the higher to the $^5E$. It is then straightforward to understand, as predicted above, that in our case the experimental peak energy separation of 0.78\,eV corresponds one-to-one to the $10Dq$ value.

It is worth looking into Fig.\,\ref{Fig5} (b) in more detail. For $10Dq$ values on the right of the gray vertical line, the ground state will no longer be the high-spin but the low-spin $3d^5$. The consequence for the $s$-NIXS spectrum is dramatic. It switches from a two-peak structure (two stars) into a five-peak feature (five stars); Fig.\,\ref{Fig10} in the Appendix explains the occurrence of these peaks. This demonstrates that the line shape of the $s$-NIXS spectrum is an extremely sensitive indicator of the ground state symmetry. The value of $10Dq$ can be determined directly from the spread of the five peaks. Consequently, the ground state hole density will also change in going from high to low-spin, \textit{i.e.} from spherical ($t_{2g}^3e_g^2$-like) to highly non-spherical ($t_{2g}^5$-like), which can be revealed directly by the image obtained from the directional dependence of the integrated $s$-NIXS intensity.

\begin{figure*}
    \centering
    \includegraphics[width=\textwidth]{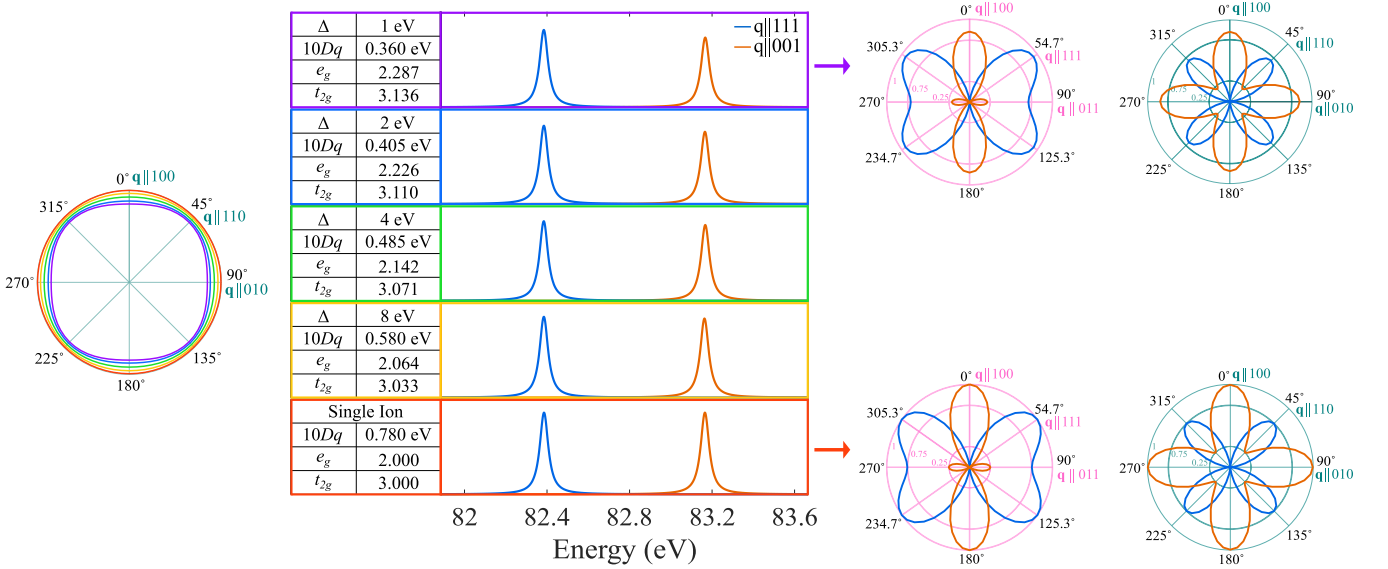}
    \caption{Mn $M_1$ simulations using a MnS$_6$ cluster calculated for various degrees of covalency. Left panel: Polar plot of the integrated intensity of the Mn $M_1$ in the [100]–[110]-[010] plane: from ionic (most outer circle) to strongly covalent (most inner distorted circle, $\Delta$ = 1 eV) case. Middle panel: Mn $M_1$ spectra along the $\mathbf{q}\parallel$ [100] (orange) and $\mathbf{q}\parallel$ [111] (blue) directions: from ionic (bottom curves) to strongly covalent (top curve, $\Delta$ = 1 eV). The $\Delta$ and corresponding ionic $10Dq$ values are indicated, together with the resulting $e_g$ and $t_{2g}$ occupation numbers. The energy splitting between the $\mathbf{q}\parallel$ [100] (orange) and $\mathbf{q}\parallel$ [111] (blue) peaks is fixed at 0.78 eV. Right panel: the directional dependence of the high (orange) and low energy peak (blue) of Mn $M_1$ spectra, for the ionic (bottom) and strongly covalent (top, $\Delta$ = 1 eV) cases.} 
    \label{Fig6}
\end{figure*}

We now investigate the influence of covalency on the $s$-NIXS image of the local $d$ hole density and the spectra. To this end, we have carried out configuration interaction calculations \cite{deGroot1994, Tanaka1994, Haverkort2012, Lu2014} using an octahedral MnS$_6$ cluster which includes explicitly the hybridization between the Mn $3d$ and the S $3p$ orbitals. We have set the hopping integrals for the $e_g$ orbital at 1.92\,eV and for the $t_{2g}$ at 1.15\,eV \cite{Haverkort2012} and varied the energy difference between the $d^5$ and $d^6\underline{L}$ configurations (charge transfer energy $\Delta$). Here $\underline{L}$ denotes the S $3p$ ligand hole states. The results are shown in Fig.\,\ref{Fig6}.

Starting with the ionic calculation, we have for the ground state an electron occupation of 3.000 for the $t_{2g}$ orbital and 2.000 for the $e_g$ (values listed in the central panel). The corresponding ground state hole density is spherical (left panel, red line), and the directional dependence of the low and high energy peaks in the calculated $s$-NIXS spectra follow the $t_{2g}$ (blue) and $e_g$ (orange) orbital shapes, respectively, as we have seen already in Fig.\,\ref{Fig2}. Switching on the hybridization between the $d^5$ and $d^6\underline{L}$ configurations, we can see that the electron occupation in the ground state increases with lowering the $\Delta$ values. It increases faster for the $e_g$ than for the $t_{2g}$ (values listed in the central panel), consistent with the fact that the hopping integral with the ligand is larger for the $e_g$ than for the $t_{2g}$. In the strongly covalent case of $\Delta=1$\,eV we have 3.136 in the $t_{2g}$ and 2.287 for the $e_g$. Accordingly, the ground state hole density, proportional to the NIXS signal, decreases for lower $\Delta$ values and becomes strongly non-spherical (left panel). This also means that the amount of hybridization can be extracted from the precise shape of the hole density as measured by $s$-NIXS.

Perhaps a surprising result is that the presence of hybridization does not have much influence on the $s$-NIXS spectrum, even for the $\Delta= 1$\,eV case. It shows the same two peak structure, and the directional dependence of the low and high energy peaks still follows the $t_{2g}$ (blue) and $e_g$ (orange) orbital shapes, respectively. It may seem surprising that hybridization or covalency does little to the spectrum, but we can draw a parallel to x-ray absorption spectrosopy (XAS), which is also a core-level spectroscopy in which a core electron is excited into the valence shell. It is known that the $M_{4,5}$ edges of Ce and the $L_{2,3}$ of the $3d$ transition metal ions can be well reproduced using ionic calculations despite the fact that there is covalency. The reason is that the energy orderings of the electron conﬁgurations are identical in the initial state and in the XAS ﬁnal state. Thus the spectral weights of the other local conﬁgurations are strongly suppressed due to quantum mechanical interference effects \cite{Gunnarsson1983,deGroot1994}. Therefore, both XAS and NIXS, generally produce a spectrum that is very similar to the one that belongs to the main local configuration, e.g. the ionic configuration. Despite the fact that the overall s-NIXS spectrum is rather insensitive to hybridization, a closer look at the line shape and the intensities does reveal details that contain information about the hybridization strength. The middle and right panels of Fig.\,\ref{Fig6} show that the intensity of the high energy peak ($e_g$, orange) becomes smaller relative to that of the low energy peak ($t_{2g}$, blue) with decreasing $\Delta$ values. The $s$-NIXS spectrum can therefore be used to help determine quantitatively the parameter values describing the hybridization process.

In the ionic calculations, we have used a $10Dq$ value of 0.78\,eV in order to get a separation of 0.78\,eV between the two peaks in the Mn $M_1$ NIXS spectrum. We will name this the ionic-$10Dq$ parameter. Upon switching on the hybridization, we must decrease the value of the ionic-$10Dq$ parameter in our calculations to mantain the 0.78\,eV separation between the two peaks in the NIXS spectrum. A larger decrease is required when $\Delta$ gets smaller, i.e. when the hybrization gets stronger (see the values in the cental panel). This can be understood if one considers the fact that the hopping integral with the ligand is larger for $e_g$ than for $t_{2g}$, and that the resulting difference in hybridization energy contributes to the energy splitting between the $e_g$ bonding state and the $t_{2g}$ equivalent. It is the combined effect of hybridization and the ionic-$10Dq$ that produces the 0.78\,eV splitting in the NIXS spectrum, which we can define as the effective-$10Dq$. NIXS, like XAS, can thus provide direct access to the effective crystal field energy \cite{deGroot1994,Tanaka1994,Agrestini2015,Wang2017}. The horizontal axis of the Sugano-Tanabe-Kamimura diagram presented in Fig. 5 can therefore be understood as the energy scale for the effective-$10Dq$ in covalent materials.

\begin{figure}
    \centering
    \includegraphics[width=\columnwidth]{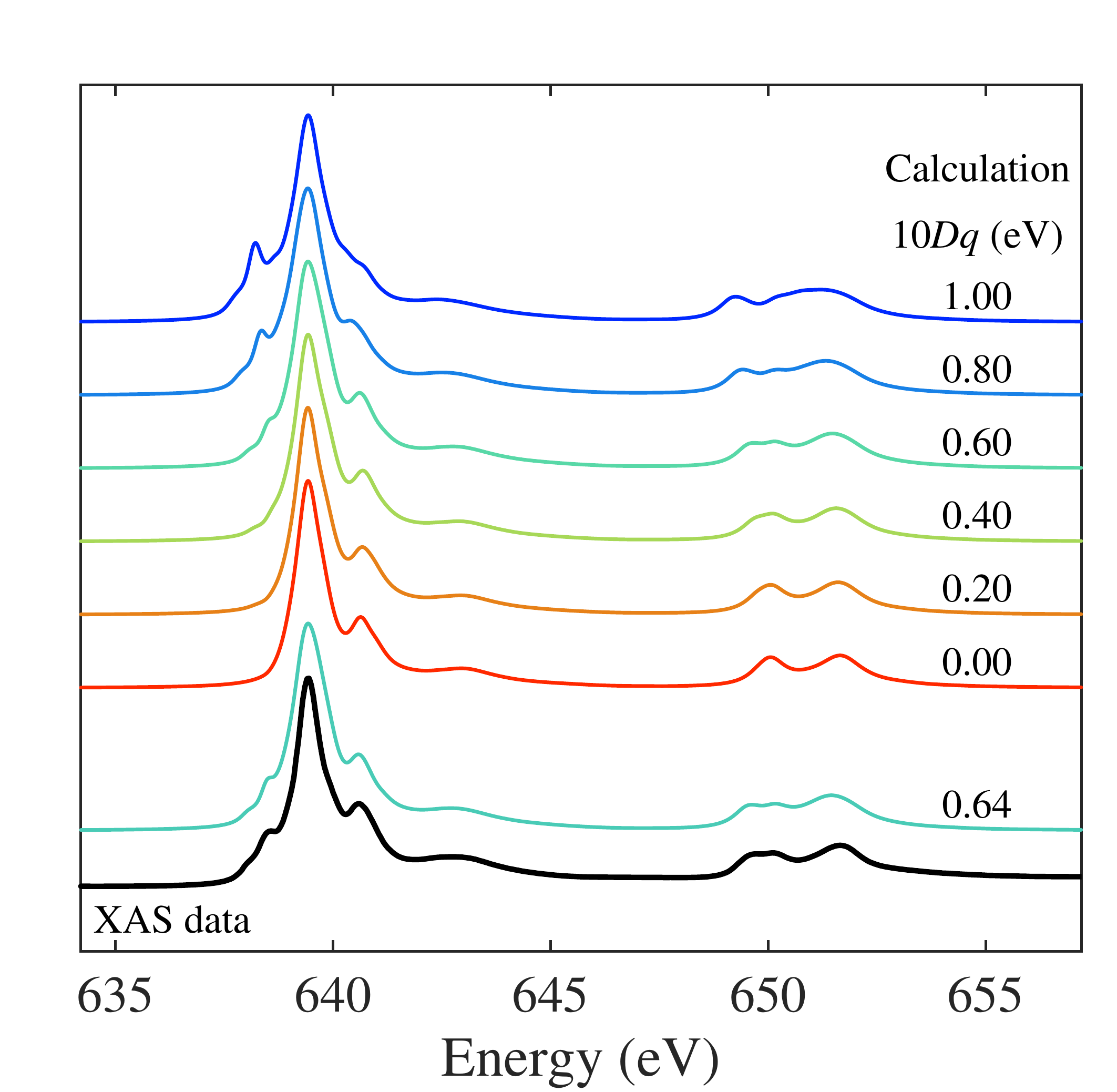}
    \caption{Mn $L_{2,3}$ x-ray absorption spectrum of $\alpha$-MnS (black line) and simulations for different values of 10$Dq$. The best fit is obtained for 10$Dq$=0.64\,eV (turquoise line). }
    \label{Fig7}
\end{figure}

To check the reliability of our analysis, we have also performed x-ray absorption spectroscopy (XAS) on the Mn $L_{2,3}$ edges, on the 11A Dragon beamline of the Taiwan Light Source at the National Syncrotron Radiation Research Center (NSRRC). The spectrum was acquired at room temperature in the total electron yield mode with a photon energy resolution of about 250\,meV.  The sample was cleaved \textit{in-situ} in a vacuum of $\approx10^{-10}$ mbar to ensure a clean sample surface. The spectrum is displayed in Fig.\,\ref{Fig7}. We have carried out ionic calculations using \textit{Quanty} \cite{Haverkort2012, Lu2014} to simulate the spectrum. The reduction of the Slater integrals was optimized in order to best reproduce the spectrum \cite{XASparameters}. 

The octahedral crystal field splitting parameter $10Dq$ is varied between -1\,eV and +1\,eV, and the best fit to the experimental spectrum was obtained for $10Dq = +0.64$\,eV. The $10Dq$ values that we have obtained from the $s$-NIXS and $L_{2,3}$ XAS measurements 
are quite close, lending confidence to the validity of our analysis. Yet, it is also worth noting that there is a difference in the values of the two experiments. We would like to infer that this has a physical origin, namely that the influence of the $2p$ core hole on the $3d$ electrons differs from that of the $3s$ core hole. An indication that this might be indeed the case comes from the observation that the Hartree-Fock value for the $F^2_{3d-3d}$ Slater-integral for the $2p^53d^6$ configuration is different from that of the $3s^13d^6$, namely (before reduction)  11.15\,eV vs. 10.53\,eV, signaling that the radial extent of the $3d$ electrons is smaller in the presence of a deep $2p$ core hole than with a more shallow $3s$. This in turn then explains why the crystal field felt by the $3d$ electrons is smaller if measured in the presence of the $2p$ core hole instead of the $3s$. Similar remarks can be made when comparing crystal field values extracted from spectroscopies with or without a core hole in the final state
\cite{Jorgensen1966,Gunnarsson1988,Cramer1991,Yasuhisa2007,Haverkort2007,Haverkort2012,Wang2017}. 

In conclusion, we have shown that $s$-NIXS has the unique ability to directly provide a spatial image of also the local excited states of transition metal ions, thereby simplifying the identiﬁcation of the multiplet character of those states. We also have shown that the well-established Sugano-Tanabe-Kamimura diagrams can be used for the analysis of the spectra since the s core hole does not add an extra orbital angular momentum. The present study thus demonstrates that one can extract from the analysis of s-NIXS spectra (1) the character of the excited states, (2) the relevant energy parameters, and (3) the character of the ground state. In addition, the integrated $s$-NIXS also gives (4) direct information about the character of the ground state. Information (4) can then be used to check the information obtained from (3). Alternatively, in case the spectra are extremely complex, information (4) can be used as a constraint for the analysis of the spectra so that (1) and (2) can be extracted more reliably. 
Therefore $s$-NIXS opens up new opportunities to determine the local electronic structure in a wide range of transition metal compounds.

\begin{figure}
    \centering
    \includegraphics[width=1\columnwidth]{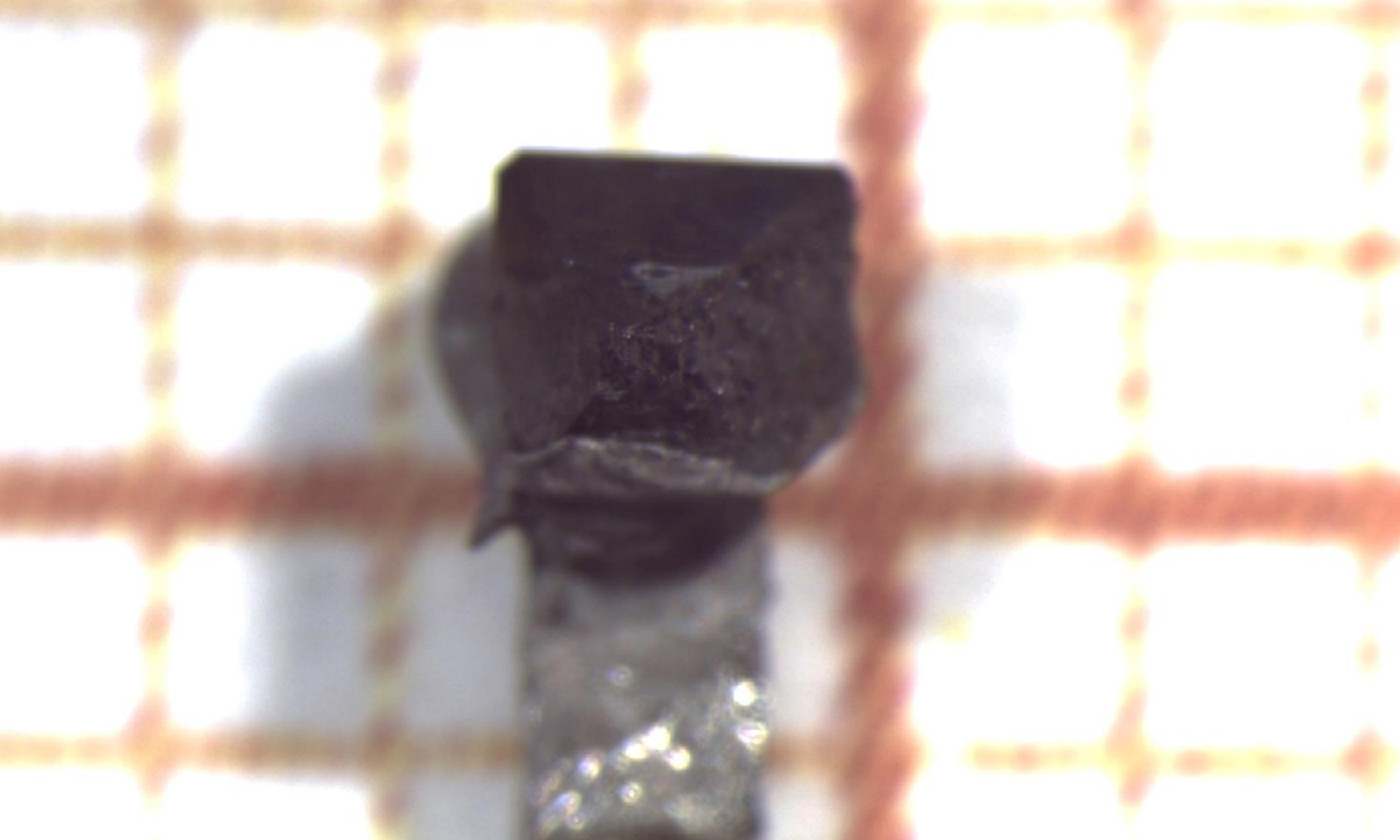}
    \caption{Top view of the $\alpha$-MnS sample (black) glued onto a stainless steel pin.}
    \label{Fig8}
\end{figure}

\begin{figure*}
    \centering
    \includegraphics[width=0.49\textwidth]{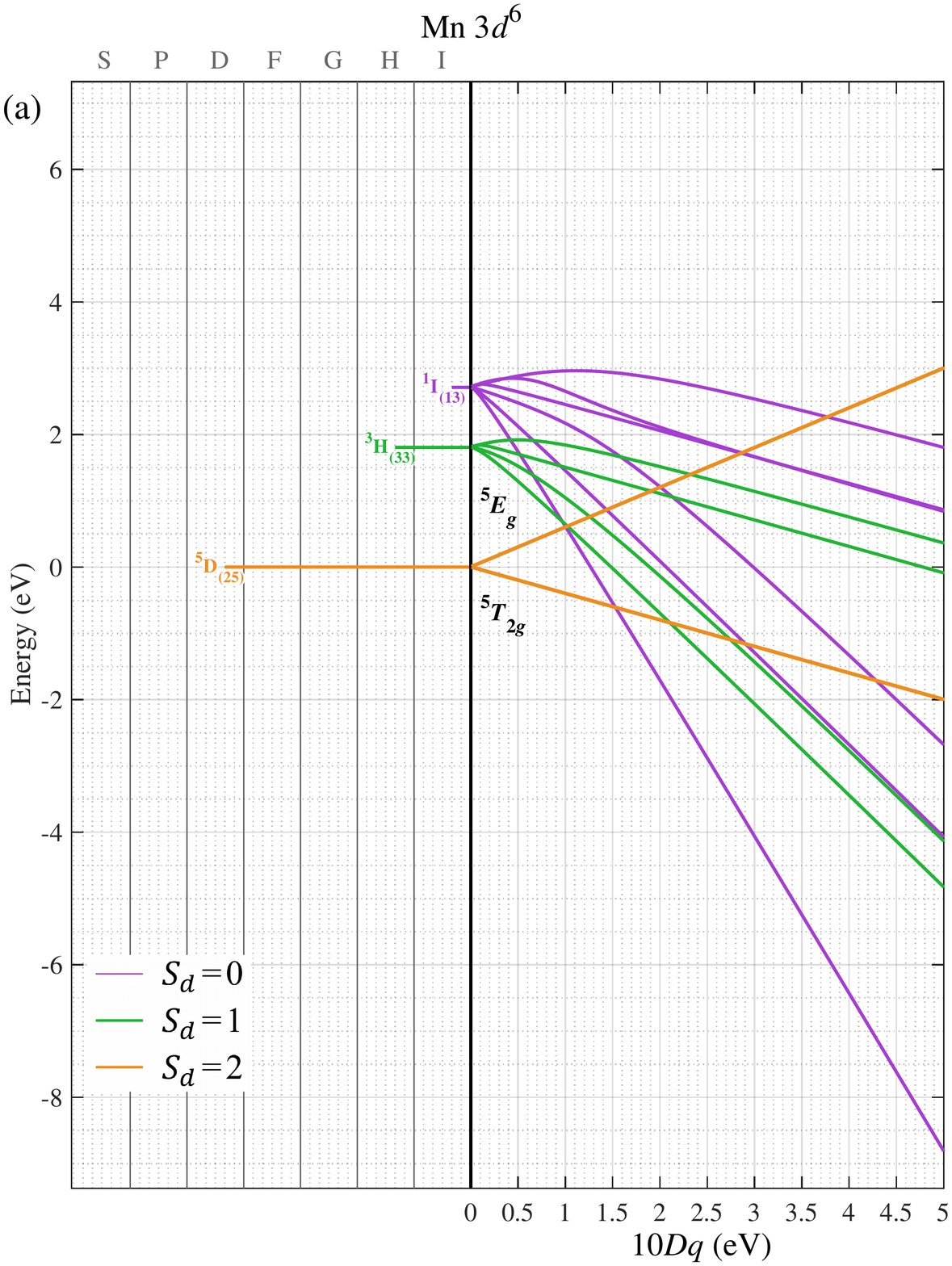}
    \includegraphics[width=0.49\textwidth]{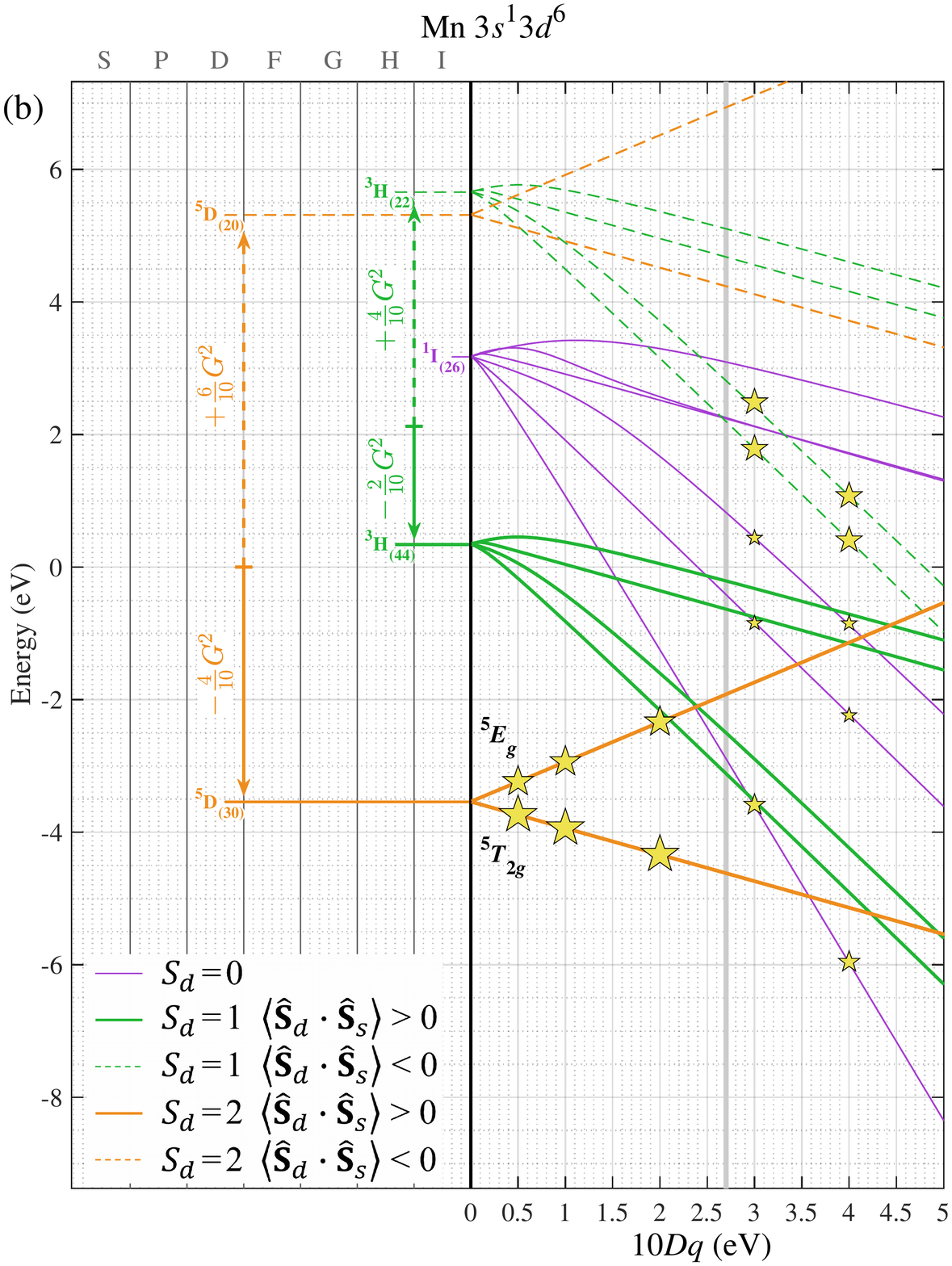}
\caption{Modified Sugano-Tanabe-Kamimura diagrams as those represented in Fig.\,\ref{Fig5}, here showing only the multiplets relevant for NIXS spectroscopy. The vertical arrows in panel (b) indicate the rigid shift in energy of the states due to the interaction with the $3s$ core hole.} 
\label{Fig9}
\end{figure*}

\section{Appendix}

Single crystals of $\alpha$-MnS were grown via chemical transport reaction using iodine as transport agent. The starting material was synthesized by direct reaction of the elements manganese (powder 99.8\% Alfa Aesar) and sulfur (pieces 99.99\% Alfa Aesar) at 975\,$^\circ$C in evacuated fused silica tubes with an inset of glassy carbon for 7 days. The obtained microcrystalline powder was recrystallized via chemical transport reaction in a temperature gradient from 1000 $^\circ$C (source) to 800\,$^\circ$C (sink). The transport agent iodine (Alfa Aesar 99,998\%) was introduced at a concentration of 8.5\,mg/cm$^3$ of the fused silica ampoule. After 14 days, the experiment was stopped by quenching the ampoule in cold water. The obtained crystals showed a well defined octahedral shape or a truncated octahedral shape depending on the growth location inside the ampoule. 
The largest crystals have edges of approximately  1\,mm length. A photograph of the sample measured is
displayed in Fig. \ref{Fig8}. Few selected small crystallites were ground and the obtained X-ray powder pattern clearly confirmed the rock salt crystal structure of $\alpha$-MnS with a lattice constant of 5.224\,\AA. \\

For more clarity on the relation between the Sugano-Tanabe-Kamimura diagrams in absence and in presence of the $3s$ core hole, the diagrams of Fig.\,\ref{Fig5} are reproduced in Fig.\,\ref{Fig9} only for the multiplets relevant for the analysis of the NIXS spectra. Figure\,\ref{Fig9} (a) shows the evolution with $10Dq$ of a singlet (purple), triplet (green) and quintet (orange) of the Mn $3d^6$ configuration. Figure\,\ref{Fig9} (b) shows the corresponding states of the Mn $3s^13d^6$ configuration (i.e. the states characterized by the same $L_d$ and $S_d$). The triplets and quintets are split into two replicas, one with parallel $\bold{S}_d$ and $\bold{S}_s$ (thick full lines) and the other with $\bold{S}_d$ anti-parallel to $\bold{S}_s$ (dashed lines). The arrows indicate the shift in energy of the replicas. The singlets (purple thin lines) are not affected by the presence of the $3s$ hole. The minor energy shift between the energy of the singlet states in the two panels is due to the different calculated values of the $F^2_{3d-3d}$ and $F^4_{3d-3d}$ Slater integrals for the two configurations. This is also the cause for the small difference between the energy of the triplet state at $10Dq=0$\,eV  in Fig.\,\ref{Fig9} (a) and the reference energy for the splitting of the corresponding replicas (starting point of the vertical green arrows) in Fig.\,\ref{Fig9} (b). Since the Slater integrals always have to be finely tuned to fit the experimental data, these minor effects do not affect the analysis of NIXS spectra and are shown here only for the sake of completeness.

\begin{figure}
    \centering
    \includegraphics[width=\columnwidth]{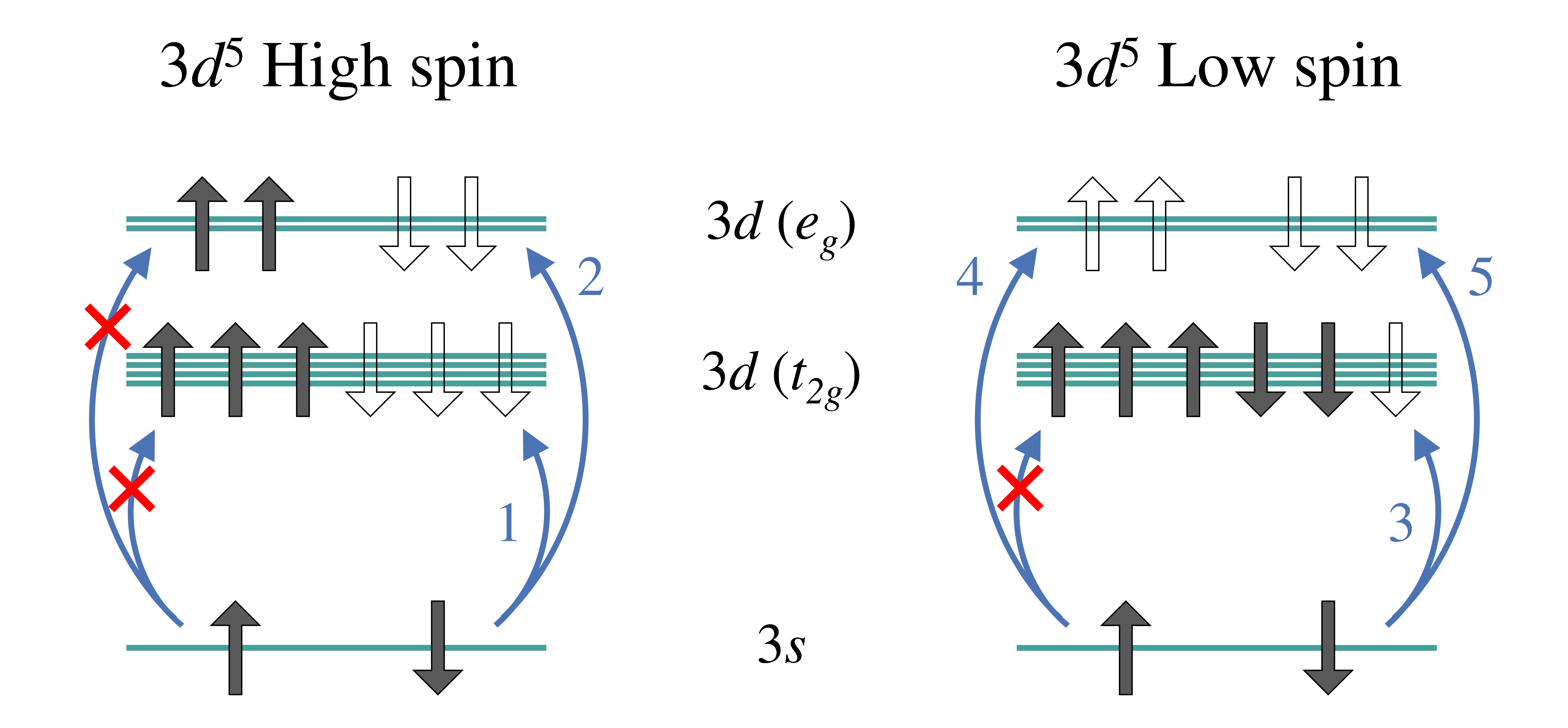}
    \caption{Schematic diagram of the $s$-NIXS excitation process for a $3d^5$ system. }
    \label{Fig10}
\end{figure}

Figure\,\ref{Fig10} shows a schematic diagram of the $s$-NIXS excitation process for a $3d^5$ system. The left panel displays the system in the high-spin state, and the right panel in the low-spin state. From the high-spin ground state, it is only possible to excite a $3s$ electron with the spin opposite to that of the half-filled $3d$ shell. This leads to ﬁnal states with $S_d = 2$ (quintet in the 3d shell) and $\braket{\hat{\bold{S}}_d\cdot\hat{\bold{S}}_s}>0$ (parallel $3s$ and $3d$ spins), corresponding to the solid orange lines in Figs.\,\ref{Fig5} (b) and \ref{Fig9} (b). There are two possibilities to do this, namely to excite the electron into a $t_{2g}$ (arrow 1 in Fig.\,\ref{Fig10}) or into an $e_g$ (arrow 2 in Fig.\,\ref{Fig10}) orbital, giving rise to two peaks in the spectrum as indicated by the two stars in Figs.\,\ref{Fig5} (b) and \ref{Fig9} (b). 

From the low-spin ground state, the excitation of the $3s$ electron to the $3d$ shell leads to $S_d = 1$ and $\braket{\hat{\bold{S}}_d\cdot\hat{\bold{S}}_s}<0$ final states (triplet in the $3d$ shell), dashed green lines in Figs.\,\ref{Fig5} (b) and \ref{Fig9} (b), or $S_d = 1$ final states (singlet in the $3d$ shell), purple lines in Figs.\,\ref{Fig5} (b) and \ref{Fig9} (b). In terms of orbital degrees of freedom, an excitation into a $t_{2g}$ orbital is only possible (arrow 3 in Fig.\,\ref{Fig10}) for the $3s$ electron having the spin opposite to the $3d$ shell, producing the low-spin $3s^13d(t_{2g})^6$ configuration. This excitation is marked by the star lying on the purple line with the steepest downward-slope in Figs.\,\ref{Fig5} (b) and \ref{Fig9} (b).

An excitation into an $e_g$ orbital, will produce a $3s^13d(e_g)^1(t_{2g})^5$ state, which can be a spin singlet or triplet as far as the intra $3d$ shell configuration is concerned. This gives rise to two peaks. These two peaks will have an intensity ratio of 1\,:\,3 reflecting the degeneracies of the spin singlet vs. triplet states. This can be understood in more detail as follows. An excitation of the $3s$ electron with the spin parallel to that of the $3d$ shell (arrow 4 in Fig.\,\ref{Fig10}) reaches a state with more up spins than down spins in the $3d$ shell, i.e. this is a spin  triplet state. On the other hand, an excitation of the $3s$ electron with the spin opposite to that of the $3d$ shell (arrow 5 in Fig.\,\ref{Fig10}) results in a state with equal amounts of up and down spins. This state belongs half to the spin singlet state and half to the triplet. The excitation into an $e_g$ orbital thus yields $\frac{1}{2}$\,:\,$\frac{3}{2}$ or 1\,:\,3 intensity ratio for the spin singlet vs. triplet states. Next we need to consider the orbital aspect of the Coulomb interactions. In this case, the Coulomb attraction between the $e_g$ electron and the $t_{2g}$ hole (in the otherwise full $t_{2g}$ subshell) depends on their relative orientations (the $x^2$-$y^2$ electron and $xy$ hole versus $3z^2$-$r^2$ electron and $xy$ hole). This orbital degree of freedom will then split each of those spin singlet/triplet states further into two. So the excitation to an $e_g$ orbital results in two small stars on the purple lines (spin singlet in the $3d$ shell) and two large stars on the dashed green lines (spin triplet in the $3d$ shell) as indicated in Figs.\,\ref{Fig5} (b) and \ref{Fig9} (b).

\section{Acknowledgments}

A.A., M.S., and A.S. acknowledge the financial support from the German funding agency the Deutsche Forschungsgemeinschaft (DFG) under Grants No SE1441-4-1 and SE1441-5-1. B.L. acknowledges support from the Max Planck-University of British Columbia Centre for Quantum Materials. The s-NIXS experiment was performed at PETRA-III at DESY, a member of the Helmholtz Association (HGF). The x-ray absorption data were acquired at the National Syncrotron Radiation Research Center (NSRRC) with the support of the Max Planck-POSTECH-Hsinchu Center for Complex Phase Materials. 
We thank C. Becker, K. H{\"o}fer, and T. Mende from MPI-CPfS, and F.-U. Dill, S. Mayer, and H. C. Wille, from PETRA-III at DESY for their skillful technical support.
\\

%\bibliography{LiteratureMnSNIXS_NoSuppl}

\end{document}